\documentclass[%
 reprint,
 amsmath,amssymb,
 aps,
]{revtex4-1}

\usepackage{graphicx}
\usepackage{dcolumn}
\usepackage{bm}
\usepackage{algorithm}
\usepackage{algorithmic}

\begin{document}

\preprint{APS/123-QED}

\title{Improvements about processing of measurement results in quantum repeater}
\thanks{A footnote to the article title}%

\author{Kenichiro Furuta}
 \email{kenichiro.furuta@toshiba.co.jp}
\affiliation{%
Corporate Research and Development Center, Toshiba Corporation
}%

\begin{abstract}
In quantum repeater protocols, measurements are performed in multiple relay points. 
Classical communications are used to convey the measurement results. 
It is important to improve the way of conveying the measurement results so that 
EPR pairs are not affected by noises during classical communications.  
One solution is execution in the blind mode\cite{blind}. 
In the blind mode, correction operations of multiple ES operations and multiple 
EPP operations are performed in a lump after EPR pairs are shared between the sender and the receiver. 
That is, the following quantum operations are performed without performing 
correction operations of the former quantum operations. 
The protocol in the conventional blind mode\cite{blind} uses concatenated CSS codes. 
Therefore, the number of qubits gets large as the communication distance gets long. 

This paper proposes quantum repeater protocols which can be executed in the blind mode 
and can be constructed by using CSS codes of single encoding. 
This result shows that posterior processing can reduce the size of encoded states in some cases. 
Additionally, the results of this paper show that posterior processing can improve performance 
of error corrections in some cases. 
The proposed protocol with single encoding can correct large errors by setting some assumptions. 
In contrast, blind mode with single encoding can deal with small errors 
if propositions in this paper are not used.  
The key ideas in posterior processing are decomposition of errors and computations for comparing 
EPR pairs of different length. 
\end{abstract}

\pacs{Valid PACS appear here}
\maketitle

\section{Introduction}
Quantum repeater protocols\cite{repeater,atomens,hbsin,hbatom,faulttol,blind,qrenc,multimem,multidim,
clusternj,allphoton,measureba} are proposed as methods for generating long EPR pairs. 
Entanglement Swapping(ES) and Entanglement Purification Protocol(EPP)\cite{epp}\cite{simpleproof} are 
key components of quantum repeater protocols. 
Generated EPR pairs can be used for the quantum teleportation\cite{teleport}. 

In order to realize quantum repeater protocols in realistic noises, various approaches are 
taken\cite{repeater,atomens,hbsin,hbatom,faulttol,blind,qrenc,multimem,multidim,clusternj,allphoton,measureba}.
Some papers try to build abstract protocols without depending on 
specific materials\cite{repeater,blind,qrenc,multimem,multidim,clusternj,allphoton,measureba}.
Among them, some methods\cite{blind}\cite{allphoton} could achieve small waiting times. 
That is, all quantum operations can be performed without waiting for arrival of measurement results of the 
former operations. 
One of the protocols which could achieve small waiting times is blind mode execution of quantum repeater protocols\cite{blind}. 

A viewpoint of the way of classifying abstract protocols is considering parts of improvements in each protocol. 
Some protocols\cite{clusternj,allphoton,measureba} concentrate focus on the way of encoding 
so that a small number of measurements on specially encoded states lead to changes 
which correspond to multiple ES operations and EPP operations. 
For example, measurements on cluster states\cite{cluster} are used in \cite{allphoton}. 
Some protocols\cite{blind} concentrate focus on the way of processing of measurement results. 
In the blind mode\cite{blind}, it is shown that correction operations which reflect measurement 
results can be performed later by revising posterior processing of measurement results. 
This paper studies the protocols which mainly focus on posterior processing.

This paper mainly focuses on posterior processing of measurement results as well as \cite{blind} 
and improve the way of posterior processing in \cite{blind}. 
The main purpose of this paper is not building quantum repeater protocols as an overall protocol, 
but proposing the efficient way of posterior processing of measurement results. 
In quantum repeater protocols, measurements are performed in multiple relay points. 
Besides, quantum states after measurements correspond to measurement results. 
So, we have to perform correction operations which reflect 
probabilistic measurement results in multiple relay points. 
Classical communications are used for conveying measurement results. 
An important task in building quantum repeater protocols is conveying measurement results 
by avoiding effects of noises on quantum states during waiting time of classical communications. 
One solution is to perform correction operations of multiple ES and EPP in a lump 
after EPR pairs are shared between the sender and the receiver.
The quantum repeater protocol based on these directions is blind mode\cite{blind}. 
The protocol in \cite{blind} uses concatenated CSS codes. 
Therefore, the number of qubits gets large as the communication distance 
gets long. 
This paper proposes quantum repeater protocols which can be executed in the blind mode 
and can be constructed by using CSS codes of single encoding. 
Additionally, the proposed protocol with single encoding can correct large errors by setting some assumptions. 
In contrast, the simple blind mode with single encoding can deal with small errors if 
propositions in this paper are not used. 
The relationship between the proposed methods and the conventional methods can be 
summarized as follows.
\begin{itemize}
\item Proposed method:\\
This method can correct large errors with single encoding by using the proposition about posterior processing.
\item Conventional blind mode with single encoding:\\
This method can correct only small errors. 
\item Conventional blind mode with concatenated codes:\\
Although this method can correct large errors, this method needs a large number of qubits.  
\end{itemize}
Therefore, the results of this paper show that posterior processing can improve performance of error corrections 
in comparison with conventional blind mode with single encoding. 
In other words, the results of this paper show that posterior processing can reduce the size of encoded state 
in comparison with conventional blind mode with concatenated coding.
The key ideas in posterior processing are decomposition of errors and computations for comparing 
EPR pairs of different length, as described in the followings. 

\section{Quantum repeater}
In this section, we explain the quantum repeater protocol which uses CSS codes in 
EPP operations\cite{simpleproof}. 
The paper of the blind mode\cite{blind} seems to use CSS codes in EPP operations.
Since this section does not consider the blind mode, corrections operations in ES and EPP 
are performed before the next ES operations and EPP operations are performed. 
We have to note that, in the case of not using blind mode, we can use CSS codes of single encoding since errors 
are corrected frequently by using classical communications during protocols and errors are not accumulated. 
In contrast, in the case of the blind mode, we have to use concatenated codes 
if we do not use special posterior processing which are proposed in this paper. 

\subsection{Whole figure of the protocol}
In the quantum repeater protocol, operations of ES are used for connecting EPR pairs and 
operations of EPP are used for increasing fidelity of EPR pairs. 
Then, long EPR pairs of high fidelity are generated. 
We assume that two EPR pairs are connected by one-time execution of ES.
Let $N-1$ be the number of relay points. 
Let $\gamma$ be the number of executions of ES operation and EPP operation.
Then, these parameters satisfy the relationship of $N=2^{\gamma}$.
Let $x$ be the order of execution of ES and EPP.
That is, ES which is executed first is called ES at $x$=1 
and EPP which is executed first is called EPP at $x$=1.
ES which is executed last is called ES at $x$=$\gamma$ 
and EPP which is executed last is called EPP at $x$=$\gamma$.
There are $N-1$ relay points between the sender $A$ and the receiver $B$, which are 
represented as $C_{1},C_{2},\cdots,C_{N-1}$.
$A$ and $B$ are represented as $C_{0}$ and $C_{N}$, respectively. 

\begin{itemize}
\item In each relay point, $C_{i}$ ($i=0,\cdots,$ $N-1$), EPR pairs are generated.
We have to note that $n$ EPR pairs are generated at each relay point 
since CSS codes of length $n$ are used with single encoding.
\item One side of each EPR pair is sent to the adjacent relay points $C_{i+1}$.
These transmissions are called quantum state transmissions.
\item From $x=1$ to $x=\gamma$, repeat the following operations.
\begin{itemize}
\item Operations of ES are performed on relay points, $C_{k \times 2^{x-1}}$ ($k=1,2,\cdots,N/2^{x-1}-1$)   
except $C_{2^{x}}$, $C_{2 \times 2^{x}}$, $\cdots,C_{N-2^{x}}$. 
\item By using classical communications, measurement results are sent to both edges of 
EPR pairs which are generated by the ES operation.
\item Correction operations which reflect measurement results are performed on both edges of 
EPR pairs which are generated by ES operations.
These operations lead to generation of EPR pairs which are 
shared between $C_{k \times 2^{x}}$ and $C_{(k+1)2^{x}}$ for $k=0,\cdots,N/2^{x}-1$.
\item Operations of EPP are performed between $C_{k \times 2^{x}}$ and $C_{(k+1)2^{x}}$ 
for $k=0,\cdots,N/2^{x}-1$. These operations lead to EPR pairs of higher fidelity.
In EPP operations, measurement operators of CSS codes are used for outputting syndromes\cite{simpleproof}. 
Since errors are corrected at every EPP operations and errors are not accumulated, 
we can use CSS codes of single encoding at every EPP operations. 
\item By using classical communications, measurement results at one side of EPR pairs 
are sent to another side of EPR pairs. 
\item Correction operations which reflect measurement results are performed on both edges of 
EPR pairs which are targets of the EPP operations.
\end{itemize}
\end{itemize}
Executing this protocol leads to the generation of EPR pairs which are shared between $C_{0}$ and $C_{N}$.

\subsection{Necessary communications}
Quantum communications are needed when EPR pairs are shared between adjacent relay points.
These quantum communications are called quantum state transmissions. 
Classical communications are needed for ES and EPP in all $x$. 
We consider classical communications which are used to share measurement 
results. Therefore, starting points of classical communications are 
relay points where measurement results are obtained. 
Destinations of classical communications are relay points where measurement results are used. 
We have to note that the destinations of communications vary with 
depending on the operational mode (blind mode execution or ordinary execution) 
of quantum repeater protocols.  

\section{Blind mode execution}
For keeping high fidelity of EPR pairs which are generated 
by quantum repeater ptrotocols, it is important to 
decrease effects of noises during waiting time. 
The method of execution in the blind mode is proposed 
as a method for decreasing effects of noises during waiting time. 

In the quantum repeater protocol, ES operations and EPP operations are performed. 
In both of ES operations and EPP operations, we can get desired EPR pairs by performing 
correction operations which correspond to measurement results.
Therefore, correction operations which correspond to measurement results are needed. 
In blind mode execution\cite{blind}, 
the next operation is done without performing correction operations of former operations.
In ES, next operations are done before performing correction operations of ES.
In EPP, next operations are done before performing correction operations of EPP. 
Correction operations in ES and EPP are done in a lump after EPR pairs are shared among the sender and the receiver.
Thus, we do not have to wait for arrival of classical communications which are used for correction operations. 

Execution in the blind mode is possible when 
quantum states without correction operations are near to one of Bell states. 
We can continue quantum repeater if targets of operations are near to one of Bell states. 
Transitions among Bell pairs are one of four Pauli operators.
Therefore, we can execute correction operations in a lump.
Correction operations at later executions are same as skipped operations. 

Blind mode execution in \cite{blind} assumes that codes used for EPP are concatenated CSS codes. 
This is because protocols are continued without correcting errors and errors are accumulated in the blind mode. 
Therefore, concatenated CSS codes are used for correcting accumulated errors. 
When CSS codes with single encoding are used without adopting posterior processing in this paper, 
accumulated errors go over the error correction capability of single code.

\section{Proposition}
This section shows our propositions. 
Execution in blind mode means that 
correction operations are performed after EPR pairs are shared
between the sender and the receiver.

In this algorithm, codes which are used for EPP are CSS codes 
as in the conventional blind mode\cite{blind}.
However, the difference is concatenation of CSS codes. 
Our proposed methods use EPP of single encoding by CSS codes in contrast to 
conventional methods which use concatenated CSS codes.

First, notations are introduced for explaining the proposed methods. 
Let $I,\sigma_{x},\sigma_{z},\sigma_{x}\sigma_{z}$ be Pauli operators, which are defined in Appendix A. 
$\sigma_{x}$ corresponds to an operator of bit errors and 
$\sigma_{z}$ corresponds an operator of phase errors. 
Let $|\phi^{+}\rangle=\frac{1}{\sqrt{2}}(|0\rangle|0\rangle+|1\rangle|1\rangle)$, 
$|\phi^{-}\rangle=\frac{1}{\sqrt{2}}(|0\rangle|0\rangle-|1\rangle|1\rangle)$, 
$|\psi^{+}\rangle=\frac{1}{\sqrt{2}}(|0\rangle|1\rangle+|1\rangle|0\rangle)$,
$|\psi^{-}\rangle=\frac{1}{\sqrt{2}}(|0\rangle|1\rangle-|1\rangle|0\rangle)$. 
These four states are called Bell states.
Correction procedures in CSS codes consist of bit error corrections and phase error corrections. 
Let $H_{1}$ be the parity check matrix which is used for bit error correction and 
$H_{2}$ be the parity check matrix which is used for phase error correction. 
Since methods for bit error corrections and phase error corrections are similar, 
we describe two corrections at the same time by representing parity matrix as $H$. 
Therefore, $H$ is either of $H_{1}$ or $H_{2}$. 
Let $e_{i,ES(j)}$ be base shifts due to Bell measurements which occur on $i$th qubit at $j$th ES.  
Let $\hat{e}_{ES(j)}$ be a vector of base shifts for all qubits which occured at $j$th ES.  
Let $\hat{s}_{EPP(j)}$ be a vector of syndromes which are obtained at $j$th EPP. 
A value 1 in error vectors means that an error occurred in the qubit. 
A value 0 in error vectors means that an error did not occur in the qubit. 
Syndromes are calculated by multiplying parity check matrices $H$ by error operators. 
We have to note that syndromes are mapped to error operators by one-to-one 
mapping if the error operators go under the error correction capability. 
Let $\sigma_{x,k}^{a_{i,j}}$ be operators of bit errors on $i$th qubit at $j$th EPP 
at the location $k$ and $\sigma_{z,k}^{b_{i,j}}$ be operators of bit errors on $i$th qubit 
at $j$th EPP at the location $k$. 
That is, a bit error is caused on $i$th qubit if $a_{i,j}=1$ and 
a phase error is caused on $i$th particle if $b_{i,j}=1$. 
Here, $i$ is the order of qubits within qubits of the same syndrome measurement. 
That is, $i$ runs from 1 to $n$ (code length).
Let $k$ indicates the location of relay point and the location inside the relay point. 
Let $|\rangle_{k}$ be quantum states at the $k$th location. 
For example, as used in Eq. (\ref{eq:bellr}), $k=1$ indicates the location of 
the sender and $k=4$ indicates the location of the receiver 
and $k=2$ and $k=3$ indicate the relay point in the middle of the sender and the receiver.
Additionally, $k=2$ indicates the left side of the middle relay point and $k=3$ indicates the right 
side of the middle relay point. 
We have to note that $\hat{e}_{ES(j)}$ and $\hat{s}_{EPP(j)}$ are computed for the 
longest EPR pairs which are shared between the sender and receiver by integrating measurement 
results for multiple sections. The way of integrating is described in the proposed method 
in this paper.   

\subsection{Strategy for blind mode with single encoding}
In blind mode, ES operations and EPP operations are performed without waiting for correction 
operations of the former operations. 
Therefore, errors of multiple ES operations and mulitple EPP operations are accumulated at the last EPP. 
For quantum repeater protocols of large distance, we cannot estimate large errors by using 
only syndromes at the last EPP since errors at the last EPP go over the error correction capability. 
Therefore, we cannot correct errors with simple methods. 
In this paper, we estimate errors at the last EPP by using syndromes of the former EPP operations. 
We estimate by assuming that the final errors are accumulation of errors between EPP 
operations of adjacent $x$. 
Additionally, we assume that each error during EPP operations of adjacent $x$ is under the error correction capability. 
With these assumptions, we can estimate each error during EPP operations of 
adjacent $x$ correctly. 
Besides, we can estimate errors at the last EPP as accumulation of each error during 
EPP operations of adjacent $x$. 
Since syndromes of CSS codes can be calculated as $|H \cdot e\rangle$, we can derive 
$|H \cdot (e_{a}+e_{b})\rangle-|H \cdot e_{a}\rangle=|H \cdot e_{b}\rangle$, 
where $e,e_{a},e_{b}$ be errors which correspond to the type of H.
This equation holds even when the weight of $e_{a}$ goes over the error correction capability. 
We can estimate $e_{b}$ from syndromes if the weight of $e_{b}$ goes under the error correction 
capability. It is not necessary that the weight of $e_{a}$ goes under the error correction capabiity. 
This characteristic is true not only for bit error corrections but also for phase error corrections. 
In the case of phase error corrections, different parity check matrix is used as shown in Appendix A. 

\begin{figure}
\includegraphics[width=9cm, height=6.5cm, bb=0 0 750 550]{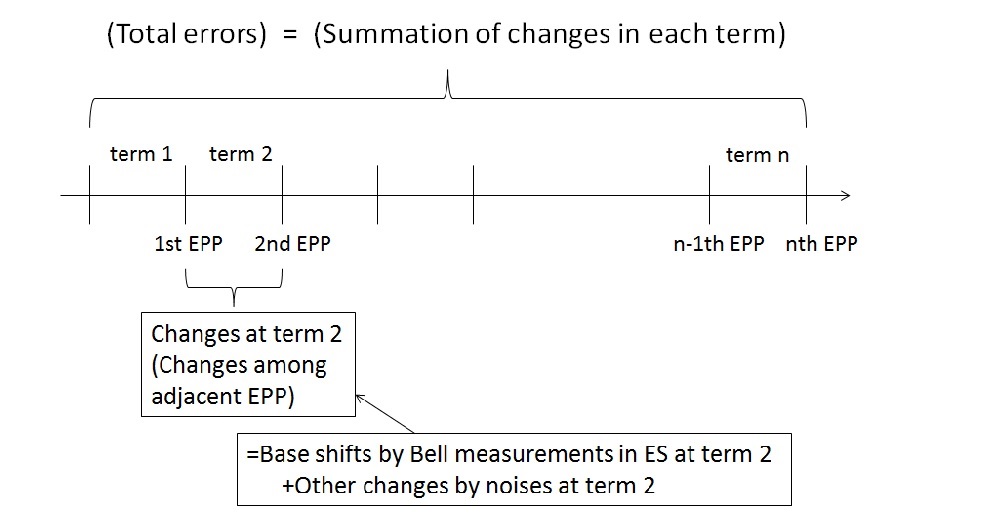}
\caption{Decomposition of errors}
\end{figure}

This characteristic about syndromes can be used for decomposing errors as the followings. 
Since syndromes of CSS codes can be calculated as $|H \cdot e\rangle$ (as explained in Appendix A), we can derive 
$|H \cdot (e_{a}+e_{b})\rangle-|H \cdot e_{a}\rangle=|H \cdot e_{b}\rangle$. 
This equation holds even when the weight of $e_{a}$ goes over the error correction capability. 
We can estimate $e_{b}$ from syndromes if the weight of $e_{b}$ goes under the error correction 
capability. It is not necessary that the weight of $e_{a}$ goes under the error correction capabiity. 

By applying this principle to estimation in quantum repater protocols, procedures can be 
described as the followings. 
For quantum repater protocols in blind mode, errors are accumulated since corrections 
are not performed until EPR pairs are shared between the sender and the receiver. 
For each error during adjacent EPP operations, error operators are $H^{-1}$(($\hat{s}_{EPP(j)}$-$H \cdot \hat{e}_{ES(j)}$)-($\hat{s}_{EPP(j-1)}$)). 
By using these operators, accumulated errors can be decomposed as Eq. (\ref{eq:dceq}).
\begin{eqnarray}
&H^{-1}(\hat{s}_{EPP(\gamma)})& \notag \\
&=H^{-1}&((\hat{s}_{EPP(\gamma)}-H \cdot \hat{e}_{ES(\gamma)})-(\hat{s}_{EPP(\gamma-1)})) \notag \\
&+H^{-1}&((\hat{s}_{EPP(\gamma-1)}-H \cdot \hat{e}_{ES(\gamma-1)})-(\hat{s}_{EPP(\gamma-2)})) \notag \\
&+&\cdots \notag \\
&+H^{-1}&((\hat{s}_{EPP(3)}-H \cdot \hat{e}_{ES(3)})-(\hat{s}_{EPP(2)})) \notag \\
&+H^{-1}&((\hat{s}_{EPP(2)}-H \cdot \hat{e}_{ES(2)})-(\hat{s}_{EPP(1)})) \notag \\
&+H^{-1}&((\hat{s}_{EPP(1)}-H \cdot \hat{e}_{ES(1)})-(\hat{s}_{EPP(0)})) \notag \\
&+&(\hat{e}_{ES(\gamma)}+\hat{e}_{ES(\gamma-1)}+\cdots+\hat{e}_{ES(1)}).
\label{eq:dceq}
\end{eqnarray}
We have to note that $\hat{s}_{EPP(0)}$ is zero vector since there are no errors at the initial time.
As we can see from this equation, overall errors can be described as summation of 
errors during adjacent EPP operations and base shifts due to ES operations as in FIG. 1. 
We have to note that this decomposition is possible since syndromes are measured 
at all EPP operations and Bell measurements are performed at all ES operations. 
By assuming that the weight of error operators of $H^{-1}$(($\hat{s}_{EPP(j)}$-$H \cdot \hat{e}_{ES(j)}$)-($\hat{s}_{EPP(j-1)}$)) 
goes under the error correction capability, we can estimate an appropriate error for each of the intervals. 
Since measurement results of Bell measurements in ES can be obtained without depending on 
syndrome measurements in EPP, the number of qubits which are affected by base shifts in ES 
is not problem. 
Estimating an appropriate error for each of the intervals and estimating base shifts due to ES operations 
lead to estimating overall errors by using the equation above. 

\begin{figure}
\includegraphics[width=10cm, height=7cm, bb=0 0 750 550]{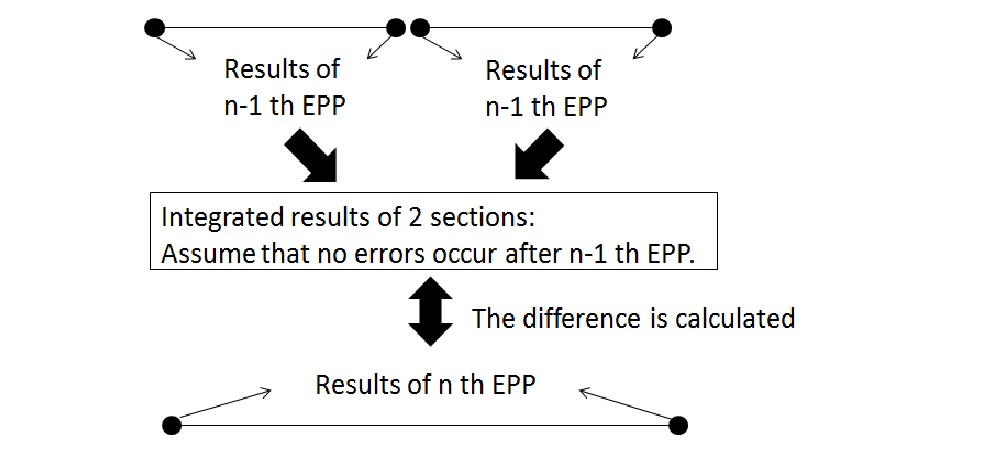}
\caption{Adjustments for comparisons}
\end{figure}

\begin{figure}
\includegraphics[width=9cm, height=6.5cm, bb=0 0 750 550]{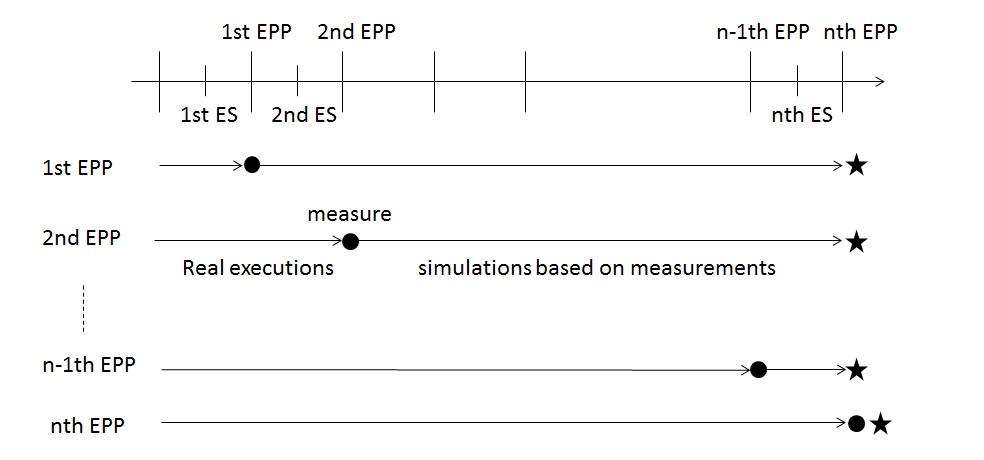}
\caption{Real execution and simulation}
\end{figure}

We have to note that decompositions above are calculated at the points of the sender and the receiver. 
Additionally, some EPR pairs stay on different sections. 
Therefore, we have to make arrangements of EPR pairs in order to 
make comparisons at the points of the sender and the receiver.  
We explain about the case of comparing EPP at $x=i$ and EPP at $x=j$, where $i<j$. 
Since the length of EPR pairs are different between $x=i$ and $x=j$, 
locations where measurement results are obtained are different. 
The measurement results in EPP at $x=i$ are processed by ES and EPP until 
$x$ reaches $j$. This processing is performed by assuming that 
no errors occur after EPP at $x=i$. We have to note that this processing 
can be performed with simulations in classical computers by using measurement 
results of EPP at $x=i$.
After measurement results at $x=i$ are processed by simulations and values at $x=j$ 
are calculated, we can compare two values since both values are for $x=j$.
FIG. 2 shows an example of comparing EPP at $x=n-1$ and EPP at $x=n$. 
FIG. 3 shows the way of comparing measurement results of EPP at all x 
by processing each of the measurement results until each $x$ reaches $n$. 
In FIG. 3, for each EPP, measurements are performed at the point of the circle and measurement results are 
processed by simulations from the point of the circle to the point of the star. 
Thus, all measurement results of EPP at different $x$ can be compared as the longest EPR pairs. 

For integrating base shifts in ES operations into one base shift which corresponds to 
an EPR pair shared between the sender and the receiver, we use the same methods. 
We integrate results of base shifts for $x$th ES operations in multiple sections 
by executing quantum repeater protocols with assuming that no errors occur 
after $x$th ES operations. 

\subsection{Outline of the proposed algorithm}
The proposed algorithm sets some assumptions about noises. 
These assumptions are fair in comparing between the proposed methods and 
the conventional methods, which is the blind mode execution with concatenated coding.  
This paper takes channel noises and memory noises into consideration. 
Additionally, this paper assumes that noises in ES operations and EPP operations and measurements 
lead only to degradation of fidelity. 
However, this paper does not take some failures into consideration, such as photon losses 
and unexecution of operations and delay of operations. 
Estimations in this paper are calculated under these assumptions. 

First, we describe the procedure transmission stage of the algorithm. 
In this stage, we use classical communications for conveying the procedure of 
the quantum repeater protocol. 
It is not problem to use classical communications in this stage since EPR pairs are 
not generated in this stage and we do not have to care about effects of quantum 
memory noise. 
\begin{itemize}
\item The sender decides the procedure of the quantum repeater protocol. 
The procedure consists of multiple commands. 
Each command consists of an ID of a relay point, a time and an operation at the time on the relay point. 
The procedure tells each relay point to execute the indicated operation at the indicated time. 
\item The sender sends the procedure to all relay points. 
\end{itemize} 
Each of the relay points execute the indicated operations at the indicated time 
without knowing situations of other relay points. 
Therefore, classical communications are not needed between the time of generating 
EPR pairs and the time of finishing sharing EPR pairs. 

After the procedure is shared among all relay points, 
ES operations and EPP operations are executed as the following. 
Since CSS codes of single encoding are used in this protocol, 
ES operations at all $x$ and EPP operations at all $x$ are 
performed on $n$ EPR pairs. Here, $n$ is the code length of CSS codes of single encoding. 
\begin{itemize}
\item In each relay point, $C_{i}$ ($i=0,\cdots,$ $N-1$), $n$ Bell states are generated, 
where $n$ is the code length of a CSS code of single encoding. 
An example of the states are $|\phi^{+}\rangle^{\otimes n}$. 
(If the required code is different from natural Bell states, Bell states are encoded 
by measuring with measurement operators of a CSS code as in \cite{simpleproof}. 
These measurements lead to convergence into an eigenspace (see Appendix A). 
Transformations into another eigenspace are performed if specific eigenspace is required.) 
\item Then, one side of each EPR pair is sent from $C_{i}$ to $C_{i+1}$. 
\item From $x=1$ to $x=\gamma$, repeat the following operations. 
ES and EPP are performed right after the preceding operation is finished. 
That is, operations are performed without waiting for arrivals of classical communications. 
Since this algorithm considers blind mode, only measurements are performed at this stage 
and correction operations which correspond to measurement results are performed at the later stage.
\begin{itemize}
\item Bell measurements of ES are performed on relay 
points, $C_{k \times 2^{x-1}}$ ($k=1,2,\cdots,N/2^{x-1}-1$) 
except $C_{2^{x}}$, $C_{2 \times 2^{x}}$, $\cdots,C_{N-2^{x}}$. 
These operations lead to generation of EPR pairs which are 
shared between $C_{k \times 2^{x}}$ and $C_{(k+1)2^{x}}$ for $k=0,\cdots,N/2^{x}-1$.
\item Measurement results at relay points where ES are performed are sent to the sender and the receiver. 
\item Syndrome measurements of EPP are performed at edges of EPR pairs, 
$C_{k \times 2^{x}}$ and $C_{(k+1)2^{x}}$ for $k=0,\cdots,N/2^{x}-1$. 
Operators of syndrome measurements at these EPP operations correspond to those of CSS codes as in \cite{simpleproof}. 
\item Measurement results at relay points where EPP are performed are sent to the sender and the receiver.
\end{itemize}
\end{itemize}

After all ES operations and all EPP operations are finished, we estimate correction operations 
based on measurement results as the followings.  
The detail of calculation of correction operations are shown in the next subsection. 
This subsection shows summary of calculating correction operations. 
In this algorithm, we assume that all ES operations and all EPP operations are 
performed as scheduled. 
\begin{itemize}
\item The sender and the receiver wait for signals of measurement results in ES and EPP. 
\item FOR {each EPR pair}
\begin{itemize}
\item We calculate correction operations for ES operations at each $x$. We also calculate 
correction operations for EPP operations separately from those of ES operations. 
This separated calculation is done by using relationship in Eq. (\ref{eq:dceq}). 
\item Correction operations which correspond to ES operations on the EPR pair are 
calculated as the followings. This calculation corresponds to calculating the latter part of the 
right side of Eq. (\ref{eq:dceq}).
\begin{itemize}
\item For each $x$, multiple EPR pairs stay on multiple sections. 
Since correction operations are performed on EPR pairs which are shared between the sender and 
the receiver, correction operations for multiple sections have to be integrated into one correction 
operation for each $x$. 
This integration is done by using Eq. (\ref{eq:bellr}) and Eq. (\ref{eq:connection}).
\end{itemize}
\item  Correction operations which correspond to EPP operations on the EPR pair are 
calculated as the followings. This calculation corresponds to calculating the former part of 
the right side of Eq. (\ref{eq:dceq}). 
\begin{itemize}
\item For each $x$, syndromes of multiple EPR pairs are summarized into one syndrome which 
correspond to EPR pairs shared between the sender and the receiver. 
This integration can be explained as in FIG. 2 and FIG. 3.
\item For each $x$, the difference of syndromes are calculated between EPP at $x$ and 
EPP at $x+1$. Then, correction operations are estimated for each term in FIG. 1. 
We have to note that effects of base shifts in ES operations can be removed in calculating this difference 
since base shifts due to ES operations are calculated as described above. 
\item Overall errors except base shifts due to ES are estimated by accumulating all of the decomposed errors 
as in FIG. 1. 
\end{itemize}
\item For each EPR pair, we perform correction operations which are computed by procedures above. 
Correction operations in ES and correction operations in EPP can be correlated as in Eq. (\ref{eq:dceq}). 
We have to note that we compute correction operations for each of bit error corrections and 
phase error corrections. 
\end{itemize}
\item END FOR
\end{itemize}
We have to note that procedures above are performed for bit error corrections 
and phase error corrections separately. 
Parity check matrix in bit error correction is different from parity check matrix in phase error correction. 

\subsection{Calculation of correction operations}
Correction operations which are used in the algorithm above 
are calculated as the followings. 

Correction operations for ES are calculated as the followings. 
In Bell measurements for connections, measurement results which correspond to 
base shifts are obtained. 
Correction operations for ES are operators of base shifts, which are one of four Pauli operators. 
Since correction operations are performed on EPR pairs which are shared between the sender and 
the receiver, correction operations for multiple sections have to be integrated into one correction 
operation for each $x$.
The way of integrations into one correction operation is similar to that in EPP 
and can be described by applying Eq. (\ref{eq:connection}), which are explained in detail in the following.

Correction operations for EPP are calculated as the followings. 
First, we explain a calculation of the primitive case. Complex calculations can be done 
by repeating this calculation. 
Correction operations which are specific to $j$th EPP are calculated by subtracting syndromes at 
$j-1$th EPP and syndromes at $j$th ES from syndromes at $j$th EPP. 
That is, the correction operations are operators which correspond to 
($\hat{s}_{EPP(j)}$-$H \cdot \hat{e}_{ES(j)}$)-($\hat{s}_{EPP(j-1)}$). 

In short, correction operations for $j$th ES are $\hat{e}_{ES(j)}$ and correction operations 
for $j$th EPP are operations which correspond to ($\hat{s}_{EPP(j)}$-$H \cdot \hat{e}_{ES(j)}$)-($\hat{s}_{EPP(j-1)}$). 

We have to note that ($\hat{s}_{EPP(j)}$-$H \cdot \hat{e}_{ES(j)}$)-($\hat{s}_{EPP(j-1)}$) 
are calculated at both edges of EPR pairs of $j$th EPP. This can be explained as the followings. 
Connection of Bell states can be described as 
\begin{eqnarray}
&|\phi^{+}\rangle_{1,2}&|\phi^{+}\rangle_{3,4} \notag \\
&=&|\phi^{+}\rangle_{1,4}|\phi^{+}\rangle_{2,3}
+|\phi^{-}\rangle_{1,4}|\phi^{-}\rangle_{2,3} \notag \\
&+&|\psi^{+}\rangle_{1,4}|\psi^{+}\rangle_{2,3}
+|\psi^{-}\rangle_{1,4}|\psi^{-}\rangle_{2,3}.
\label{eq:bellr}
\end{eqnarray} 
Eq. (\ref{eq:bellr}) means that Bell measurements on particle 2 and 3 lead to 
convergence into one of four Bell states on particle 1 and 4. 
Besides, Bell states on particle 2 and 3 correspond to Bell states on particle 1 and 4.
Therefore, we can know which EPR pairs are obtained at particle 1 and 4 by 
watching states on particle 2 and 3. 
Performing Pauli operators on the left side of Eq. (\ref{eq:bellr}) leads to the 
following equation for each $i$ $(i=1,...,n)$. 

\begin{eqnarray}
&(&\sigma_{x,1}^{a_{i,j-1}}\sigma_{z,1}^{b_{i,j-1}}\sigma_{x,2}^{a_{i,j-1}}\sigma_{z,2}^{b_{i,j-1}})|\phi^{+}\rangle_{1,2} \notag \\
&\otimes& (\sigma_{x,3}^{a_{i,j-1}}\sigma_{z,3}^{b_{i,j-1}}\sigma_{x,4}^{a_{i,j-1}}\sigma_{z,4}^{b_{i,j-1}})
|\phi^{+}\rangle_{3,4} \notag \\
&=&(\sigma_{x,1}^{a_{i,j-1}}\sigma_{z,1}^{b_{i,j-1}}\sigma_{x,4}^{a_{i,j-1}}\sigma_{z,4}^{b_{i,j-1}} \notag \\
&\otimes& \sigma_{x,2}^{a_{i,j-1}}\sigma_{z,2}^{b_{i,j-1}}\sigma_{x,3}^{a_{i,j-1}}\sigma_{z,3}^{b_{i,j-1}}) \notag \\
&\times& (|\phi^{+}\rangle_{1,4} \otimes |\phi^{+}\rangle_{2,3}
+|\phi^{-}\rangle_{1,4} \otimes |\phi^{-}\rangle_{2,3} \notag \\
&+&|\psi^{+}\rangle_{1,4} \otimes |\psi^{+}\rangle_{2,3}
+|\psi^{-}\rangle_{1,4} \otimes |\psi^{-}\rangle_{2,3}).
\label{eq:connection}
\end{eqnarray}
The left side of Eq. (\ref{eq:connection}) means EPR pairs before connections. 
The right side of Eq. (\ref{eq:connection}) has 4 terms. 
Each of the 4 terms corresponds to EPR pairs after connection. 
$\hat{s}_{EPP(j-1)}$ means syndromes which correspond to 
$\sigma_{x,1}^{a_{i,j-1}}\sigma_{z,1}^{b_{i,j-1}}\sigma_{x,4}^{a_{i,j-1}}\sigma_{z,4}^{b_{i,j-1}}$ of all $i$. 
$\hat{e}_{ES(j)}$ means operators which change states after Bell measurements on particle 1 and 4 into 
the desired term of the right side of Eq. (\ref{eq:connection}). 
Therefore, $\hat{e}_{ES(j)}$ changes a Bell state to another Bell state. 
$\hat{s}_{EPP(j)}$ means syndromes which correspond to 
$\sigma_{x,1}^{a_{i,j}}\sigma_{z,1}^{b_{i,j}}\sigma_{x,4}^{a_{i,j}}\sigma_{z,4}^{b_{i,j}}$ of all $i$. 
When changes occur only by Bell measurements, error operators satisfy 
$\sigma_{x,1}^{a_{i,j-1}}\sigma_{z,1}^{b_{i,j-1}}\sigma_{x,4}^{a_{i,j-1}}\sigma_{z,4}^{b_{i,j-1}}$
=$\sigma_{x,1}^{a_{i,j}}\sigma_{z,1}^{b_{i,j}}\sigma_{x,4}^{a_{i,j}}\sigma_{z,4}^{b_{i,j}}$ for all $i$. 
However, due to other noises, errors are not only shift changes at ES. 
Therefore, error operators at EPP(j) can be different from error operators at EPP(j-1). 
By using Eq. (\ref{eq:connection}), we can compute correction operations for each ES and EPP. 
In other words, we can integrating errors of multiple EPP(j-1) into one errors on longer EPR pairs 
by processing quantum repeater protocols with assuming that no errors occur after EPP(j-1). 

For estimating overall errors, we summarize the decomposed errors. 
Since the longest EPR pairs in protocols are shared between the sender and the receiver, 
we compute the difference of adjacent EPP operations by integrating shorter EPR pairs 
into the longest EPR pairs, which are shared between the sender and the receiver. 
We can integrate errors for each section into 
errors of larger sections as in the calculation of the primitive case above and FIG. 2. 
Repeating these integrations lead to computing integrated errors on EPR pairs 
shared between the sender and the receiver. 
The methods of computing the differences are shown in FIG. 3. 

\section{Performance of the proposed protocols}
This section compares the difference between the conventional blind mode 
and the proposed method. 
As stated in the introduction of this paper, the main focus of this paper is posterior processing of 
measurement results. 
Therefore, it is appropriate to compare with protocols which focus on posterior processing. 
Even if some specifications are inferior to those of protocols which focus on encoding, 
it is not problem since focuses are different. 
Improvements in the proposed method may be used for posterior processing part of protocols 
which focus on encodings and can enhance specifications of protocols which focus on encodings. 
Additionally, since propositions in this paper focus on posterior processing of measurement results, 
the most important thing which we should analyze is correctness of estimating correction operations 
and conditions for the correctness. 
We suppose that correction operations are estimated after EPR pairs are shared between the sender and 
the receiver by using measurement results which are obtained before EPR pairs are shared between 
the sender and the receiver. 
Therefore, we have to check that estimation of correction operations is valid even after the following 
ES operations and EPP operations are performed. 

This paper sets some assumptions about noises. 
These assumptions are fair in comparing between the proposed methods and 
the conventional blind mode. 
This paper takes channel noises and memory noises into consideration. 
Additionally, this paper assumes that noises in ES operations and EPP operations and measurements 
lead only to degradation of fidelity. 
However, this paper does not take some failures into consideration, such as photon losses 
and unexecution of operations and delay of operations. 
Estimations in this paper are calculated under these assumptions. 

\subsection{Correctness of estimation}
Since operators for measuring syndromes in EPP correspond to those of CSS codes, 
EPP has a parameter of the error correction capability, which are denoted as $t$. 
We can specify errors as long as the weight of an error is less than $t$ since 
an error and syndromes can be mapped by one-to-one mapping in this case. 
Therefore, we can trace accumulation of errors if the weight of each error during each interval 
of syndrome measurements is less than $t$ qubits. 
This means that we can specify errors even when the weight of accumulated errors goes over $t$,  
if the weight of an error during each interval of syndrome measurements is less than $t$. 

By applying this method to quantum repeater protocols, we can correct errors of large weight 
if we measure syndromes at each EPP and if the weight of each error during adjacent EPP 
operations is less than $t$. 
This is the new assumption which is needed for blind mode with single encoding.

Additionally, since base shifts due to ES operations can be detected and corrected without dependence on errors in EPP, 
the number of qubits which are affected by base shifts due to ES is not problem. 
Therefore, correcting large errors is possible if the weight of an error which corresponds to 
($\hat{s}_{EPP(j)}$-$H \cdot \hat{e}_{ES(j)}$)-($\hat{s}_{EPP(j-1)}$) 
are less than $t$ for each j in both of bit error corrections and phase error corrections.

Correctness of this estimation can be shown by decomposing accumulated errors as follows. 
Overall errors can be decomposed by using Eq. (\ref{eq:dceq}).
As we can see from this equation, overall errors can be described as summation of 
errors during adjacent EPP operations and base shifts due to ES operations as in FIG. 1. 
By assuming that the weight of error operators of $H^{-1}$(($\hat{s}_{EPP(j)}$-$H \cdot \hat{e}_{ES(j)}$)-($\hat{s}_{EPP(j-1)}$)) 
goes under the error correction capability, we can estimate an appropriate error for each of the intervals 
since an error and an syndrome can be mapped by one-to-one mapping in this case. 
Additionally, since measurement results of Bell measurements in ES can be obtained without 
depending on syndrome measurements in EPP, the number of qubits which are affected by 
base shifts in ES is not problem. 
Estimating an appropriate error for each of the intervals and estimating base shifts due to ES operations 
lead to estimating overall errors by using Eq. (\ref{eq:dceq}). 
Discussions above show that correction operations match 
if the weight of an error between adjacent EPP operations goes under the error correction capability. 
Since the proposed protocol does not consider photon losses and unexecutions of operations, 
nonexistence of these errors are assumed. 

\subsection{Inheritance of major merits}
This subsection shows that the proposed protocol could reduce resources 
without spoiling major merits of the conventional blind mode. 
We have to note that minor points of the conventional blind mode 
are changed. That is, additional assumptions about frequency of errors are 
changed, as described in the previous section. 

What we could improve in the proposed method is the number of EPR pairs 
which are used in the protocol. 
The number of EPR pairs in the conventional method is $N$*$n^{\gamma}$. 
The number of EPR pairs in the proposed method is $N*n$. 
This is because the proposed protocol could execute by using CSS codes with 
single encoding. 

Although the proposed protocol could reduce resources, the proposed protocol 
inherits major merits of the conventional blind mode. 
\begin{itemize}
\item One of the merits in the conventional blind mode is 
smallness of waiting times. 
The expected time of executing protocols is as follows. 
The time for executing the conventional method is almost same as the time 
for executing the proposed method. 
Both of the time above are the summation of the time of delivering EPR pairs to adjacent nodes and 
the time of executing ES operations and EPP operations since these protocols are executed in the 
blind mode. 
The time for classical communications of sending measurement results are not considered since 
this time does not affect the fidelity of generated EPR pairs. 
\item The other merit of the conventional blind mode is sending small states via channels by 
using EPP in \cite{simpleproof}. For example, one side of $|\phi^{+}\rangle^{\otimes n}$ are sent via channels in the EPP.
In both of the conventional blind mode and the proposed blind mode, 
quantum states which are sent via channels can be small EPR pairs, such as isolated $|\phi^{+}\rangle$, 
by using the EPP\cite{simpleproof}. 
\item Another point of comaprison is adjustments to errors. 
The type of errors to which the proposed protocol adjust is almost same as the 
type of errors to which the conventional protocol adjust. 
The type of errors is degradation of EPR pairs in channel transmissions and in ES operations and  
noises during waiting times. 
Both of the proposed method and the conventional method could adjust to these errors as long as 
errors are within the error correction capability of entanglment purification. 
The small difference is setting additional assumptions in the proposed method. 
The assumption is that each of the errors during adjacent EPP operations goes under 
the error correction capability. The detail of this assumption is explained in the previous section. 

However, the proposed protocol and the conventional blind mode did not consider photon losses 
and unexecution of operations. 
\end{itemize}
Thus, we could show that the proposed protocol could reduce resources without spoiling merits 
of the conventional blind mode except the new assumption above. 

The results of this paper show that posterior processing of measurement results can 
reduce the size of encoded states with keeping high correction capability. 
Since one of the hurdles against realization of some quantum repeater protocols is 
largeness of encoded states, essence in this paper will be useful for reducing the size 
of the encoded states by improving posterior processing of measurement results. 

\section{conclusion}
We constructed concrete methods of blind mode with single encoding. 
Our proposed methods do not use concatenated CSS codes, 
but use single encoding by CSS codes. 
We showed that the proposed method can correct large errors 
by using CSS codes of single encoding 
if overall errors are decomposed and each of the decomposed errors 
goes under the error correction capability. 
The results of this paper show that posterior processing can improve performance of error corrections 
in comparison with the conventional blind mode with single encoding. 
The proposed protocol with single encoding can correct large errors by setting some assumptions. 
In contrast, blind mode with single encoding can deal with small errors 
if propositions in this paper are not used. 
In other words, the results of this paper show that posterior processing can reduce the size of 
encoded states in comparison with the conventional blind mode with concatenated codes. 
In summary, the proposed way about posterior processing is effective for improving performance of 
error corrections and for redusing the size of encoded states. 

%
%
%

\appendix
\section{CSS codes and EPP}
In this section, we explain CSS codes\cite{css} and its conversion to EPP\cite{simpleproof}\cite{generalepp}. 

\subsection{Code construction and error representation}
Since CSS codes are subclass of stabilizer codes, 
CSS codes inherit characteristics of stabilizer codes. 
First, I describe stabilizer codes. Then, I describe CSS codes 
as special cases of stabilizer codes. 
Possible errors can be represented as superposition of some error bases. 
These bases belong to a group, which is called as the error group. 
In order to construct the error group, we define Pauli operators as in Eq. (\ref{eq:Pauli}). 

\begin{eqnarray}
I&=&\left[ \begin{array}{ccc}1&0\\0&1
\end{array} \right],
\sigma_{x}=\left[ \begin{array}{ccc}0&1\\1&0
\end{array} \right], \notag \\
\sigma_{z}&=&\left[ \begin{array}{ccc}1&0\\0&-1
\end{array} \right],
\sigma_{x}\sigma_{z}=\left[ \begin{array}{ccc}0&-1\\1&0
\end{array} \right].
\label{eq:Pauli}
\end{eqnarray}
By using Eq. (\ref{eq:Pauli}), the error group can be written as 
\begin{eqnarray}
E^{n}= \{ \pm M_{1} \otimes M_{2} \otimes \cdots \otimes M_{n} \mid \notag \\
M_{1},\cdots,M_{n} \in \{ I,\sigma_{x},\sigma_{z},\sigma_{x}\sigma_{z} \} \}. 
\label{eq:errorgroup}
\end{eqnarray}
We describe elements of the error group as 
$E_{i}^{n}=E_{i}^{n|1} \otimes E_{i}^{n|2} \otimes \cdots \otimes E_{i}^{n|n}$, 
where $E_{i}^{n|1},\cdots,E_{i}^{n|n} $ are one of 4 Pauli operators. 
Codewords of stabilizer codes are elements of eigenspaces of subgroups of the error group. 
Let a subgroup of the error group be $P$. Codewords are elements of an eigenspace of $P$. 
An eigenspace means a space of states which have the same eigenvalue for all operators in $P$. 

There are some methods for representing quantum errors such as 
methods by state vectors and methods by density operators. 
In this paper, we describe the change of state vectors. 
Let $|\Psi\rangle$ be a codeword of stabilizer codes.
Quantum states of $n$ qubits change by unitary operators as the following. 
\begin{eqnarray} 
U_{n}|\Psi\rangle=\sum_{i}\left(c_{i}E_{i}^{n}|\Psi\rangle\right)= \notag \\
\sum_{i,j}c_{i,j}E_{i}^{m}|\psi\rangle \otimes E_{j}^{n-m}|\phi\rangle. \label{eq:nqbtrans} 
\end{eqnarray} 
Here, $E_{i}^{n} \in E^{n}$. 
$|\Psi\rangle$ consists of $n$ qubits. 
Let $m$ qubits out of the $n$ qubits be system and be described as $|\psi\rangle$. 
Let the other $n-m$ qubits be environment and be described as $|\phi\rangle$. 
Setting $\displaystyle \sum_{j}c_{i,j}E^{n-m}_{j} |\phi\rangle=\nu_{i}$ and 
outputting syndromes $a_{i}$ into ancilla leads to 
\begin{equation} 
U_{n}|\Psi\rangle=\sum_{i}E^{m}_{i} |\psi\rangle \otimes \nu_{i} \otimes |a_{i}\rangle. \label{eq:sysenvtrans4}
\end{equation} 
Measuring ancilla leads to 
\begin{equation} 
E^{m}_{i} |\psi\rangle \otimes \nu_{i} \otimes |a_{i}\rangle. \label{eq:sysenvtrans5}
\end{equation} 

Measuring ancilla of Eq. (\ref{eq:sysenvtrans4}) leads to conversion into Eq. (\ref{eq:sysenvtrans5}). 
Eq. (\ref{eq:sysenvtrans5}) represents codewords which are affected by one of error bases. 
Thus, measuring ancilla leads to conversion into bases of errors. 
Then, we can specify errors on codewords. 

\subsection{Error correction capability} 
We define the weight of $E_{i}^{n}$ in the following. 
Let this weight be $\gamma \left( E_{i}^{n} \right)$. 
$E_{i}^{n}$ consists of tensor products of $n$ Pauli operators (Eq. (\ref{eq:Pauli})). 
The weight of $E_{i}^{n}$ means the number of operators which are not $I$ out of $n$ operators. 
Stabilizer codes of minimum distance $\theta$ can correct error bases whose weight goes under 
$\lfloor \left( \theta -1 \right) /2 \rfloor.$ 
Thus, stabilizer codes of minimum distance 
$\theta$ can correct errors which occurred on less than $\lfloor \left( \theta -1 \right) /2 \rfloor$ qubits.
This parameter is called the error correction capability. 

\subsection{Codes construction of CSS codes}
In CSS codes, codewords $|\Psi\rangle$ are constructed as follows.
Suppose $C_{1}$ and $C_{2}$ are $[n,k_{1}]$ and $[n,k_{2}]$ classical linear codes such that $C_{2}$$\subset$$C_{1}$ 
and $C_{1}$ and $C_{2}^{\perp}$ both correct t errors. 
By using the $C_{1}$ and $C_{2}$, CSS code can be constructed as 
$|x+C_{2}\rangle$=$\frac{1}{\sqrt{|C_{2}|}}$$\displaystyle\sum_{y \in C_{2}}$$|x+y\rangle$, 
where $x \in C_{1}$ is any codeword in $C_{1}$.
Let $H_{1}$ be parity check matrix of $C_{1}$ and $H_{2}$ be parity check matrix of $C_{2}$. 
Then, syndromes of CSS codes correspond to $|H_{1} \cdot e_{1}\rangle$ for bit error corrections 
and $|H_{2} \cdot e_{2}\rangle$ for phase error corrections. 
That is, states after outputting syndromes of bit error $e_{1}$ are 
$|x+C_{2}\rangle$=$\frac{1}{\sqrt{|C_{2}|}}$$\displaystyle\sum_{y \in C_{2}}$$|x+y+e_{1}\rangle|H_{1} \cdot e_{1}\rangle$.

\subsection{Algorithms for error correction} 
In the error correction stage, we perform unitary operators to output syndromes on ancilla. 
We specify errors based on syndromes which are measured on ancilla. 
Operators to output syndromes are decided by the subgroup $P$. 
In CSS codes, these operators can be divided into two categories. 
One is for bit error detection and another is for phase error detection. 
Therefore, in CSS codes, the error correction stage can be divided into 
the bit error correction stage and the phase error correction stage. 
The phase error correction stage is executed by performing WH translations before 
measuring syndromes and correcting errors. WH translations are performed again 
after correcting phase errors. 

\subsection{Conversion to EPP} 
Concrete methods of general conversion into EPP are written in \cite{simpleproof},\cite{generalepp} 
as summarized in the followings. 
\begin{itemize}
\item Alice prepares multiple Bell states. The number of Bell states corresponds to the code length. 
(If the required code is different from natural Bell states, Alice encodes these Bell states by measuring 
with measurement operators of stabilizer codes or CSS codes.) 
\item Alice sends one side of each EPR pair to Bob. 
\item Alice and Bob measure syndromes by using measurement operators 
which correspond to those of stabilizer codes or CSS codes. 
\item Alice and Bob exchange syndrome information and specify errors 
based on syndrome information. 
\end{itemize}
As above, measurement operators in EPP correspond to those of 
stabilizer codes or CSS codes. 
Thus, EPP can correct errors when the number of error qubits goes under 
the error correction capability. 

\nocite{*}

\bibliography{reference}

\end{document}